\documentclass[%
 amsmath,amssymb,
 aps,
pre,floatfix,superscriptaddress,twocolumn,showpacs]{revtex4-1}

\usepackage{lipsum}
\usepackage{comment}
\usepackage{graphicx}
\usepackage{dcolumn}
\usepackage{bm}

\usepackage{latexsym}
\usepackage{ucs}
\usepackage[utf8]{inputenc}
\usepackage{eurosym}
\usepackage{siunitx}
\usepackage{wrapfig}
\usepackage{amsmath,amsfonts,amssymb}
\usepackage{url}
\usepackage{gensymb} 
\usepackage{natbib}
\usepackage{multirow}
\usepackage{xcolor}
\usepackage[normalem]{ulem}
\usepackage[colorlinks]{hyperref}  
\usepackage{cleveref}
\hypersetup{linkcolor=blue,citecolor=blue,urlcolor=blue}
\newcommand{\mv}[1]{\ensuremath{\mathbf{#1}}} 
\newcommand{\gv}[1]{\ensuremath{\mbox{\boldmath$ #1 $}}} 
\newcommand{\abs}[1]{\left| #1 \right|} 
\newcommand{\avg}[1]{\left \langle #1 \right \rangle} 
\newcommand{\der}[2]{\frac{d #1}{d #2}} 
\newcommand{\pd}[2]{\frac{\partial #1}{\partial #2}} 

\crefname{figure}{Fig.}{Figs.}
\Crefname{figure}{Figures}{Figures}

\begin{document}

\title{Microscopic theory for nonequilibrium correlation functions in dense active fluids}

\author{Vincent E. Debets}
\thanks{Authors contributed equally}
\affiliation{Department of Applied Physics, Eindhoven University of Technology, P.O.Box 513, 5600MB Eindhoven, The Netherlands} 
\affiliation{Institute for Complex Molecular Systems, Eindhoven University of Technology, P.O.Box 513, 5600MB Eindhoven, The Netherlands}
\author{Lila Sarfati}
\thanks{Authors contributed equally}
\affiliation{Department of Applied Physics, Eindhoven University of Technology, P.O.Box 513, 5600MB Eindhoven, The Netherlands}
\affiliation{Département de Physique de l'Ecole Normale Supérieure, ENS, Université PSL, 24 rue Lhomond, 75230, Paris Cedex 05, France}
\author{Thomas Voigtmann}
\affiliation{Institut f\"ur Materialphysik im Weltraum, Deutsches Zentrum f\"ur Luft- und Raumfahrt, 51170 K\"oln, Germany} 
\affiliation{Department of Physics, Heinrich-Heine Universit\"at D\"usseldorf, Universit\"atsstraße 1, 40225 D\"usseldorf, Germany}
\author{Liesbeth M. C. Janssen}
\email{l.m.c.janssen@tue.nl}
\affiliation{Department of Applied Physics, Eindhoven University of Technology, P.O.Box 513, 5600MB Eindhoven, The Netherlands} 
\affiliation{Institute for Complex Molecular Systems, Eindhoven University of Technology, P.O.Box 513, 5600MB Eindhoven, The Netherlands}

\date{\today}

\begin{abstract} 
One of the key hallmarks of dense active matter in the liquid, supercooled, and solid phases is so-called equal-time velocity correlations. Crucially, these correlations can emerge spontaneously, i.e., they require no explicit alignment interactions, and therefore represent a generic feature of dense active matter. This indicates that for a meaningful comparison or possible mapping between active and passive liquids one not only needs to understand their structural properties, but also the impact of these velocity correlations. This has already prompted several simulation and theoretical studies, though they are mostly focused on athermal systems and thus overlook the effect of translational diffusion. Here we present a fully microscopic method to calculate nonequilibrium correlations in systems of thermal active Brownian particles (ABPs). We use the integration through transients (ITT) formalism together with (active) mode-coupling theory (MCT) and analytically calculate qualitatively consistent static structure factors and active velocity correlations. We complement our theoretical results with simulations of both thermal and athermal ABPs which exemplify the disruptive role that thermal noise has on velocity correlations.     

\end{abstract}

\maketitle

\newcommand{\argc}[1]{\left[#1\right]} 
\newcommand{\arga}[1]{\left\lbrace #1\right\rbrace } 
\newcommand{\argp}[1]{\left(#1\right)} 
\newcommand{\valabs}[1]{\vert #1\vert} 
\newcommand{\moy}[1]{\left\langle  #1 \right\rangle } 
\newcommand{\projP}{\mathcal{P}}
\newcommand{\Adens}[3]{\rho_{#1}(\boldsymbol{#2}_{#3})} 
\newcommand{\tAdens}[3]{\tilde{\rho}_{#1}(\boldsymbol{#2}_{#3})}

\section{Introduction}
Bridging biology and physics, active matter has received particular interest over the past two decades and continues to remain at the vanguard of biophysical and soft matter research~\cite{Bechinger2016,Ramaswamy2010,Marchetti2013rev}. Comprising of particles that convert energy into systematic movement or mechanical work, active systems are intrinsically out-of-equilibrium and ubiquitous in living matter. Concurrently, synthetic active materials are also increasingly being experimentally realized providing an interesting playground for colloidal science far away from equilibrium. The appeal of studying and perhaps utilizing active systems comes from their ability to showcase a wealth of new nonequilibrium phenomena that cannot be observed in standard passive matter. Notable examples include motility induced phase separation (MIPS)~\cite{Buttinoni2013,Ginot2018,Palacci2013,Linden2019}, activity-induced crystallization~\cite{Briand2016,Ni2014}, accumulation around repulsive obstacles~\cite{Berke2008}, and active turbulence~\cite{Giomi2015,Alert2022rev}. 

Another key nonequilibrium hallmark is so-called equal-time velocity correlations~\cite{Szamel_2021}, which primarily arise in the context of dense active matter, a regime that has recently seen a significant rise of interest in part due to its implications in diseases such as cancer and asthma~\cite{Janssen2019active,Berthier2019review,Lang2018,Grosser2021}. These correlations quantify local cooperative (or aligned) particle motion and were first extracted in confluent cell monolayers~\cite{Garcia2015cell,Angelini2011cell,Angelini2010}. Since then they have also been extensively studied in simulations of self-propelled particles where they appear in, e.g., MIPS~\cite{Caprini2020}, dense active (glassy) fluids~\cite{BerthierAOUP2017,FlennerAOUP2016,Flenner2020,Caprini2020velocity1,Caprini2021inertial,Keta2022}, model cell layers~\cite{Henkes2020}, and chiral active matter~\cite{Debets2023chiral}, while they also naturally surface in mode-coupling theories (MCTs) of dense active fluids~\cite{SzamelAOUP2016,SzamelABP2019}. Importantly, these correlations have mostly been shown to emerge spontaneously, that is, they in principle do not necessitate any explicit alignment interactions, and thus represent a robust feature of any dense active matter system. 

This implies that for a meaningful comparison and perhaps a mapping between a dense active and passive fluid, one not only needs to understand the influence of activity on structural and dynamic correlations, but more crucially also understand the role of (equal-time) velocity correlations. 
To this end, several theoretical approaches have already been brought forward, though they primarily look at large length scales (or equivalently small wavenumbers)~\cite{Caprini2020,Henkes2020,Szamel_2021}. Moreover, to our knowledge most studies on velocity correlations (with Ref.~\cite{Caprini2021inertial} a notable exception) have mainly focused on athermal systems, that is, systems without translational noise, and thus do not consider the disruptive effect the latter can have on velocity correlations.

In this work, to add to our fundamental understanding of dense active matter, we present an entirely microscopic approach to analytically calculate nonequilibrium (equal-time) correlations for interacting thermal active Brownian particles (ABPs). Our method uses the integration through transients (ITT) formalism~\cite{Fuchs2009} in conjunction with active MCT~\cite{Voigtmann2017} and is applied on the static structure factor and the so-called longitudinal velocity correlations (as they hold particular relevance for active MCTs~\cite{SzamelAOUP2016,SzamelABP2019}). More specifically, we look at the individual correlations 
that comprise the total longitudinal velocity correlation function, which allows us to better pinpoint its exact origins. To complement the theoretical results and rationalize the role of thermal noise, we also perform simulations of both thermal and athermal ABPs. We demonstrate that our theory can qualitatively describe the nonequilibrium structure factor and active-active velocity correlations, while our simulations clearly illustrate the dominant effect thermal noise has on velocity correlations (especially ones involving interaction forces).

\section{Integration through transients}

Given its suitability for the ITT formalism and prevalence in active matter, we take as our model dense active liquid a collection of $N$ interacting two-dimensional (2D) ABPs (disks) at a number density $\rho=N/V$ and temperature $T$. The position $\mv{r}_{i}$ of each particle $i$ evolves in time according to~\cite{Ramaswamy2010,Bechinger2016,Voigtmann2017} 
\begin{equation}\label{eom_r}
    \der{\mv{r}_{i}}{t} = \zeta^{-1} (\mv{F}_{i} + \mv{f}_{i}) + \gv{\xi}_{i}.
\end{equation}
where $\zeta$ is the friction coefficient, $\boldsymbol{F}_i$ is the interaction force, and $\gv{\xi}_{i}$ represents a Gaussian thermal noise with zero mean and variance $\avg{\gv{\xi}_{i}(t)\gv{\xi}_{j}(t^{\prime})}_{\mathrm{noise}}=2D_{\mathrm{t}}\mv{I}\delta_{ij}\delta(t-t^{\prime})$, with $D_{\mathrm{t}}=k_{B}T\zeta^{-1}$ the diffusion coefficient and $\mv{I}$ the unit matrix. The self-propulsion speed $v_{0}$ is assumed constant so that the active force equals $\mv{f}_{i}=\zeta v_{0}\mv{e}_{i}$. The orientation of the self-propulsion velocity is in turn given by $\boldsymbol{e}_{i} = [\cos(\theta_i),\sin(\theta_i)]$ and its angle randomly reorients with a rotational diffusion coefficient $D_{\mathrm{r}}$, i.e.,
\begin{equation}\label{eom_theta}
    \dot{\theta}_{i} = \chi_{i},
\end{equation}
where $\chi_{i}$ denotes a Gaussian noise process with zero mean and variance $\avg{\chi_{i}(t)\chi_{j}(t^{\prime})}_{\mathrm{noise}}=2D_{\mathrm{r}}\delta_{ij}\delta(t-t^{\prime})$.
Based on the equations of motion one can derive the following Smoluchowski operator,
\begin{equation}\label{Scholuchowski_ABPs}
       \Omega = \sum\limits_{j=1}^N D_t\nabla_j \cdot(\nabla_j - \beta\boldsymbol{F}_j) + D_r \partial_{\theta_j}^2 - v_0 \nabla_j \cdot\boldsymbol{e}_j
\end{equation}
which governs the time-evolution of the probability distribution function (PDF) of particle positions and orientations $P(t)$ via,
\begin{equation}\label{eomP}
    \pd{P}{t} = \Omega P(t).
\end{equation}
In equilibrium, that is, for $v_0 = 0$, this equation admits a Boltzmann solution $P_{\mathrm{eq}} \propto e^{-\beta U}$ with $U$ the total potential energy from which the interaction force $\mv{F}_{j}=-\nabla_{j}U$ is derived and $\beta=(k_{B}T)^{-1}$ the inverse thermal energy.

The starting point of the ITT approach is then to employ the identity $e^{\Omega t} = 1 + \int_0^t dt' e^{\Omega t'}\Omega$ and insert it in the formal solution of the PDF, $P(t)=e^{\Omega t}P(0)$~\cite{Fuchs2009,Sharma2016,Voigtmann2017}. Letting our ABP system switch from an equilibrated passive state ($v_{0}=0$) to an active state ($v_{0}>0$) at time $t=0$ and assuming it has reached an active steady-state at $t\rightarrow \infty$, one can retrieve an exact expression for the active steady-state average of any observable $A$~\cite{Voigtmann2017},
\begin{equation}\label{ITT_ABP}
    \moy{A}_{\mathrm{ss}} = \moy{A}_{\mathrm{eq}} - \beta v_0 \int_0^\infty dt \moy{\sum\limits_{j=1}^N \boldsymbol{e}_j\cdot\boldsymbol{F}_j e^{\Omega^\dagger t}A}_{\mathrm{eq}}.
\end{equation}
Here, $-\beta v_0 \sum\limits_{j=1}^N \boldsymbol{e}_j\cdot\boldsymbol{F}_j = \Omega P_{\mathrm{eq}}/P_{\mathrm{eq}}$. $\Omega^\dagger$ represents the adjoint Smoluchowski operator, which acts on everything to its right except the PDF and is given by
\begin{equation}\label{adjoint_Schmoluchowski_ABPs}
    \Omega^\dagger = \sum\limits_{j=1}^N D_\mathrm{t}(\nabla_j + \beta\boldsymbol{F}_j)\cdot \nabla_j + D_\mathrm{r} \partial_{\theta_j}^2 + v_0  \boldsymbol{e}_j\cdot\nabla_j.
\end{equation} 
The appeal of this approach comes from the fact that a usually intractable steady-state average $\avg{\hdots}_{\mathrm{ss}}$ is now rewritten in terms of transient quantities $\avg{\hdots}_{\mathrm{eq}}$ which are averaged with respect to the known equilibrium distribution $P_{\mathrm{eq}}$. At the same time, the integral term in~\cref{ITT_ABP} remains highly nontrivial and provides a serious obstacle for any analytical progress. 

\Cref{ITT_ABP} together with the expectation that many-body effects on the stationary averages strongly couple to density fluctuations, has lead to the development of an active mode-coupling theory (MCT) for the transient time-dependent density correlation function~\cite{Voigtmann2017,Reichert2020modecoupling,Reichert2020tracer,Reichert2021rev,DebetsMCT2022}, 
\begin{equation}
    S_{ll'}(\boldsymbol{k},t) = \moy{\rho_l^*(\boldsymbol{k}) e^{\Omega^\dagger t}\rho_{l'}(\boldsymbol{k})}_{\mathrm{eq}},
\end{equation}
where $\rho_l(\boldsymbol{k}) = \frac{1}{\sqrt{N}} i^l \sum\limits_{j=1}^N e^{i \boldsymbol{k}\cdot \boldsymbol{r}_j}e^{il\theta_j}$ depicts the density mode (the factor $i^{l}$ is added for technical convenience, see~\cref{sec:real_ISF} for precise details).
Using only the passive static structure factor $S(k)$ (from for instance liquid state theory or simulations), which enters in the initial condition via,
\begin{equation}
  S_{ll'}(k) = \moy{\rho_l^*(\boldsymbol{k})\rho_{l'}(\boldsymbol{k})}_{\mathrm{eq}}=\delta_{ll'}[1+ \delta_{l0}(S(k)-1)],  
\end{equation}
one can invoke this theory to find self-consistent solutions for $S_{ll'}(\boldsymbol{k},t)$. Besides aiding to our fundamental understanding of glassy active matter~\cite{Voigtmann2017,Reichert2020modecoupling,Reichert2020tracer,Reichert2021rev,DebetsMCT2022}, these solutions in conjunction with additional MCT-approximations also provide an interesting pathway to explicit analytical expressions for complete steady-state correlation functions, i.e., ones that include the integral term in~\cref{ITT_ABP}. This has already been successfully explored in the context of the average swim velocity~\cite{Reichert2021rev,Reichert_20_Thesis}, but our aim is to generalize the idea and also apply it to arguably more complex static structure and velocity correlations. 

In particular, let us first define the integral term in \cref{ITT_ABP} for any static observable $A$ as,
\begin{equation}\label{CA_def}
    C^{A}(t)\equiv\moy{\sum\limits_{j=1}^N \boldsymbol{e}_j\cdot\boldsymbol{F}_j e^{\Omega^\dagger t}A},
\end{equation}
where, for notational convenience, we have omitted the subscript 'eq' which will be done from this point onward. In the context of rheology MCT-approximations have been successfully applied directly to $C^{A}(t)$~\cite{Fuchs2009}, even though more sophisticated approximations exist~\cite{VogelFuchs_2020}.
However, in our case $C^{A}(t)$ does not yet lend itself to MCT-approximations as it may lead to relatively large and thus nonphysical contributions from the integral term (this has also been explicitly checked for the observables in this work). To avoid this problem and normalize the integral term it has been suggested in Ref.~\cite{Reichert2021rev} to further reduce $C^{A}(t)$ by means of an irreducible time-evolution operator, $\Omega_{\text{irr}}^\dagger = \Omega^\dagger - \projP_v$, with
\begin{equation}\label{def_Pv_ITT_S}
    \projP_v =-\sum\limits_{ij}\left. |\boldsymbol{F}_i\cdot\boldsymbol{e}_i\right> \frac{\beta^2D_t}{N} \left<\boldsymbol{F}_j\cdot\boldsymbol{e}_j | \right. .
\end{equation} 
This definition is in part motivated by the fact that MCT-approximations are usually better suited for slow variables and thus we want to project out the active part of the evolution operator which is assumed to take on a fast character.
Using Dyson decomposition one then finds
\begin{equation}\label{CCirr_def}
\int_{0}^{\infty}dtC^A(t) = \frac{\int_{0}^{\infty}dtC_{\text{irr}}^A(t)}{1+ \frac{\beta^2 D_t}{N}\int_{0}^{\infty}dtC_{\text{irr}}^v(t)}
\end{equation}
where we have introduced the irreducible correlation of swim velocity corrections
\begin{equation}
    C_{\text{irr}}^v(t) = \moy{\sum\limits_{j=1}^N \boldsymbol{e}_j\cdot\boldsymbol{F}_j e^{\Omega_{\text{irr}}^\dagger t}\sum\limits_{i=1}^N \boldsymbol{e}_i\cdot\boldsymbol{F}_i},
\end{equation}
and $C_{\text{irr}}^A$, which is the same correlation function as~\cref{CA_def} only evolving in time with irreducible dynamics $e^{\Omega_{\mathrm{irr}}^{\dagger}t}$ instead of full dynamics $e^{\Omega^{\dagger}t}$. 

Employing customary MCT-approximations~\cite{Szamel1991,gotze2008complex,Nagele1999,Debets2021Brownian}, that is, two projections on density doublets, a factorization of dynamic four-point correlations into products of two-point correlations, and replacing irreducible by full dynamics, it has been shown that~\cite{Reichert2021rev} (see also~\cref{sec:analy_details} for more details),
\begin{equation}\label{Cirrv_expression}
\begin{aligned}
   & \frac{D_t\beta^2}{N}C_{\text{irr}}^v(t)= \frac{\rho D_t}{4\pi}  \int_0^\infty dq\, q^3 c(q)^2 \\
   &\times \argc{\argp{\tilde{S}_{11}(q,t)-\tilde{S}_{-11}(q,t)}\tilde{S}_{00}(q,t) + 2\tilde{S}_{01}(q,t)\tilde{S}_{10}(q,t)}.
    \end{aligned}
\end{equation}
Here, we have introduced the real quantity $\tilde{S}_{ll^\prime}(q,t)=e^{i(l-l^\prime)\theta_{q}}S_{ll^\prime}(\mv{q},t)$, which, due to rotational symmetry, is independent of the orientation of the wavevector $\theta_q$ and only depends on its magnitude $q$~\cite{Reichert2021rev,Reichert_20_Thesis} (see also~\cref{sec:real_ISF}). Note that $C_{\text{irr}}^v(t)$ is now fully written in terms of transient dynamic density correlation functions $S_{ll'}(\boldsymbol{k},t)$. Moreover, it is responsible for normalizing the integral term as its magnitude grows when one approaches denser conditions.

The final step then consists of applying the same MCT-approximations to $C_{\text{irr}}^A(t)$ to end up with an explicit expression for the steady-state average $\avg{A}_{\mathrm{ss}}$ that only depends on $S_{ll'}(\boldsymbol{k},t)$, its passive counterpart $\avg{A}$, and the relevant control parameters. As such, ITT in conjunction with active MCT provides a generic framework for analytically evaluating (static) non-equilibrium averages, though we will show that it gives the most qualitatively consistent results when our observable $A$ can already be written as a sum of density doublets.

\section{Methods}
\subsection{Active-MCT and ITT numerics}
To utilize the proposed ITT method and calculate steady-state averages, we require explicit expressions for $S_{ll^\prime}(\mv{k},t)$. We therefore numerically solve the active-MCT equations (as detailed in e.g., Refs.~\cite{Voigtmann2017,Reichert2020modecoupling,DebetsMCT2022}) for a monodisperse colloidal system of hard disks of diameter $\sigma$. We use an equidistant wavenumber grid $k\sigma=[0.3,0.5,\hdots,39.9]$ (note that we drop the smallest wavenumber $k\sigma=0.1$ in favor of numerical stability) and perform the integration over time according to the algorithm presented in Ref.~\cite{Voigtmann2017}. For the latter, we calculate the first $N_{t}/2=16$ points in time using a Taylor expansion with a step size $\Delta t=10^{-6}$, numerically integrate the equations of motion for the next $N_{t}/2$ points in time, duplicate the timestep, and repeat the process.
As input we employ an analytical expression for $S(k)$ attained (as a function of the area fraction $\phi=\rho\pi\sigma^{2}/4$) from density functional theory ~\cite{Thorneywork2018}. For computational convenience we only consider the first two non-trivial active modes $l\in [-1,0,1]$, which is sufficient to calculate the correlation functions in this work, and fix the wavevector along the $x$-axis, i.e.\ $\mv{k}=k \mv{e}_{x}$, so that $S_{ll^\prime}(\mv{k},t)=\tilde{S}_{ll^\prime}(k,t)$ is always real. We also set the area fraction at $\phi=0.6$ (though we have checked that $\phi=0.5$ gives similar results) which is a trade-off between allowing sufficiently dense conditions and numerical stability for small values of $k$ and large active speeds $v_{0}$. All results are presented in units of $\sigma$ and $\sigma^{2}/D$ for distance and time respectively. Finally, the time-integration of the ITT equations is carried out using the trapezoidal rule.

\subsection{Simulation details}
To complement our theoretical results and characterize the role of translational (thermal) noise on steady-state correlations in dense active matter, we perform simulations of a slightly polydisperse mixture of $N=1000$ quasi-hard ABPs (disks). The dynamics of each particle $i$ is governed by~\cref{eom_r,eom_theta} where the interaction force $\mv{F}_{i}=-\sum_{j \neq i} \nabla_{i} V_{\alpha\beta}(r_{ij})$ is derived from a quasi-hard-sphere powerlaw potential $V_{\alpha\beta}(r)= \epsilon\left( \frac{\sigma_{\alpha\beta}}{r}\right)^{36}$~\cite{Weysser2010structural,Lange2009}. The interaction energy $\epsilon$ and friction constant $\zeta$ are equal to one. For the thermal simulations we fix the temperature and thus the diffusion coefficient at $T=D_{\mathrm{t}}=1.0$, whereas they are strictly zero for the athermal simulations. To ensure polydispersity, our mixture consists of equal fractions of particles with diameters (in units of $\sigma$) $\sigma_{\alpha\alpha}=\{0.8495,0.9511,1.0,1.0489,1.1505\}$\footnote{Particle diameters are chosen such that the first four moments correspond to the results of a Gaussian distribution with a mean of $1$ and a standard deviation of $0.1$.}, which are additive so that $\sigma_{\alpha\beta}=(\sigma_{\alpha\alpha}+\sigma_{\beta\beta})/2$~\cite{DebetsMCT2022}. Simulations consist of solving the Langevin equation [\cref{eom_r}] via a forward Euler scheme and are carried out using LAMMPS~\cite{Lammps}. We fix the square box size to set the area fraction at $\phi=0.75$ which is slightly denser than the theoretical values. This is done to mitigate the effects of motility induced phase separation (MIPS)~\cite{Keta2022} and allow for a better comparison with the ITT results which are obtained for an assumed homogeneous system. Setting the persistence time and active speed, we then run the system for approximately $200$ time units to ensure we are in a steady state, and afterwards track the particle positions in time. 
In all simulation results, $\sigma$, $\epsilon$, and $\zeta \sigma^{2}/\epsilon$ denote the units of length, energy, and time respectively~\cite{Flenner2005}.

\section{Results \& discussion}
\subsection{Density correlations}
We begin by considering the steady-state (or nonequilibrium) static structure factor, i.e., we let $A=\rho_0^*(\boldsymbol{k})\rho_{0}(\boldsymbol{k})$, and define it as $S_{\mathrm{neq}}(k)=\avg{\rho_0^*(\boldsymbol{k})\rho_{0}(\boldsymbol{k})}_{\mathrm{ss}}$. Carrying out the MCT-approximations [where we mention that, because $A$ has the form of a density doublet, this requires one less projection on density doublets compared to $C_{\text{irr}}^v(t)$], one can find the following ITT expression (see~\cref{sec:analy_details} for more details),  
\begin{equation}\label{S_ITTexpression}
    S_{\mathrm{neq}}(k) = S(k) - 2\rho v_0 k\, c(k)
    \frac{\int_0^\infty dt \, \tilde{S}_{10}(k,t)\tilde{S}_{00}(k,t)}
    { 1 + \frac{D_t\beta^2}{N}\int_0^\infty dt\, C_{\text{irr}}^v(t)}.
\end{equation}
Note that it only depends on the magnitude $k$ which is consistent with the fact that the active steady-state remains isotropic. We also point out that the time-integrals in~\cref{S_ITTexpression} consist of products containing at least one of the correlation functions $\tilde{S}_{ll^\prime}(k,t)$ that always decay to zero, even in the ideal glass state~\cite{Voigtmann2017}. $S_{\mathrm{neq}}(k)$ thus never diverges which is to be expected. Interestingly, this feature also holds for all other static correlation functions discussed in this work.

\begin{figure}[ht!]
    \centering
    \includegraphics[width=0.48\textwidth]{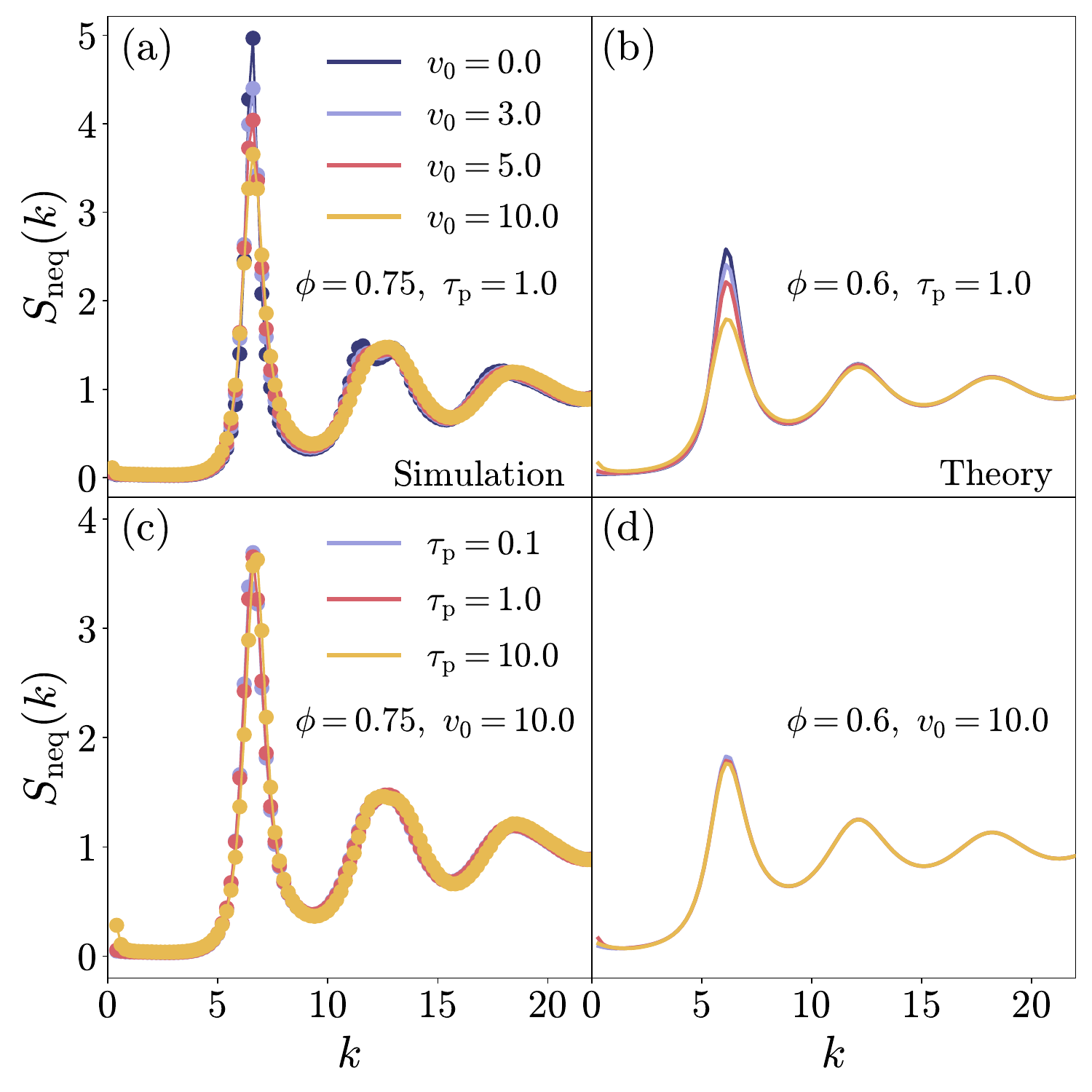}
    \caption{The steady-state structure factor $S_{\mathrm{neq}}(k)$ as a function of wavenumber $k$ directly measured from simulation data of thermal ABPs (a,c) or obtained fully analytically via the ITT formalism (b,d). Results correspond to different active speeds (a-b) and persistence times (c-d).}
    \label{Sk_thermal}
\end{figure}

Based on the above equation we have calculated $S_{\mathrm{neq}}(k)$ for different active speeds $v_{0}$ and persistence times $\tau_{\mathrm{p}}=D_{\mathrm{r}}^{-1}$. Concomitantly, we have extracted $S_{\mathrm{neq}}(k)$ from simulation data for the same active control parameters. The results are shown in~\cref{Sk_thermal} and upon first glance look qualitatively consistent. In particular, the location of the first peak remains almost constant at $k\sim2\pi/\sigma$, while its height decreases significantly for increasing values of $v_{0}$ and only marginally for increasing $\tau_{\mathrm{p}}$ (note that in simulations the peak is higher due to the larger packing fraction $\phi$). These features are well captured by the theoretical predictions and are likely a result of the (quasi-)hard nature of the particle interactions (constant first peak location) and the system exhibiting faster dynamics upon increased activity (decreasing first peak height). 
Beyond the first minimum, which is lifted slightly upwards, the influence of activity on $S_{\mathrm{neq}}(k)$ becomes fairly small. The only exception is the splitting of the second peak at $v_{0}=0$ (a marker for local crystalline order~\cite{vandewaal1999} and thus not captured by the theory), which disappears when the system becomes more active and thus more fluid. 

At large enough persistence and active speed we also observe a small increase of $S_{\mathrm{neq}}(k)$ when $k$ approaches zero which is associated with increased compressibility and usually interpreted as a precursor for MIPS~\cite{Biniossek2018}. In that regard, it is remarkable that the ITT result is able to pick up on this as in principle it has no notion of phase separation. 
Finally, we have also extracted $S_{\mathrm{neq}}(k)$ for athermal systems (see \cref{Sk_omegak_active_athermal}), but their qualitative features are mostly similar to those of their thermal counterparts (assuming the system is not undergoing MIPS which dramatically changes the structure factor).

\subsection{Velocity correlations}
Having started from a purely structural correlation, a natural next step is now to also try to explicitly connect the structure to the (active) velocity of individual particles. For that, we take a closer look at the so-called (longitudinal) velocity correlation function which is defined as,  
\begin{equation}\label{def_vv}
    \omega(k)  = \frac{1}{N\zeta^2} 
     \hat{\boldsymbol{k}}\cdot\moy{\sum\limits_{i=1}^{N}\argp{\boldsymbol{F}_i + \boldsymbol{f}_i}e^{-i\boldsymbol{k}\cdot\boldsymbol{r}_i}\sum\limits_{j=1}^{N} \argp{\boldsymbol{F}_j + \boldsymbol{f}_j}e^{i\boldsymbol{k}\cdot\boldsymbol{r}_j}}_{\mathrm{ss}}\cdot\hat{\boldsymbol{k}},
\end{equation}
and has already been frequently studied in the context of athermal dense active matter where it is shown to only depend on the magnitude $k$, develop oscillations upon increasing $v_{0}$ and $\tau_{\mathrm{p}}$, and become constant in the passive limit~\cite{FlennerAOUP2016,BerthierAOUP2017,Flenner2020,Szamel_2021}. Its behavior for a thermal system, which includes translational diffusion, in contrast remains largely unexplored and our aim is therefore to characterize this velocity correlation function in thermal systems and make a comparison with the better-known athermal phenomenology.

Since it turns out that our theoretical approach is better equipped to deal with specific terms in~\cref{def_vv} and to more exactly pinpoint the role of thermal noise on this correlation function, we have decided to separate the velocity correlations in three distinct contributions and consider these individually. As such, we write
\begin{equation}
    \omega(k)=\omega_{\mathrm{a}}(k)+2\omega_{\mathrm{c}}(k)+\omega_{\mathrm{int}}(k),
\end{equation}
where $\omega_{\mathrm{a}}(k)$ represents the coupling between the active forces (or equivalently active velocities), $\omega_{\mathrm{c}}(k)$ the cross correlation between active and interaction forces, and $\omega_{\mathrm{int}}(k)$ the correlation between interaction forces. From this point onward, we will refer to these terms as the active-active, active-passive, and passive-passive velocity correlations respectively.



\subsubsection{Active-active velocity correlations}
Employing MCT-approximations one can derive an explicit ITT expression for the active-active velocity correlation function which is given by (see~\cref{omega_a_der} for a precise derivation),
\begin{equation}\label{vv_ffexpression_ITT}
    \omega_{\mathrm{a}}(k) = \frac{v_0^2}{2} - v_0^3 \rho k c(k)
    \frac{\int_0^\infty dt \tilde{S}_{01}(k,t)\big[\tilde{S}_{11}(k,t)-\tilde{S}_{-11}(k,t)\big]}
    { 1 +  \frac{D_t \beta^2}{N}\int_0^\infty dt \, C_{\text{irr}}^v(t)}.
\end{equation}
It is worth pointing out that the observable associated with this contribution can again be rewritten in terms of density doublets [see~\cref{active_densitymode}] and thus, similar to $S_\mathrm{neq}(k)$, one less projection on density doublets is required. Moreover, within the ITT approximation $\omega_{\mathrm{a}}(k)$ only depends on the magnitude of the wavevector $k$ and not its orientation. This is in agreement with the behavior found for $\omega(k)$ in simulations and is to be expected if the active steady-state is isotropic~\cite{FlennerAOUP2016,BerthierAOUP2017,Flenner2020,Szamel_2021}.

\begin{figure}[ht!]
    \centering
    \includegraphics[width=0.48\textwidth]{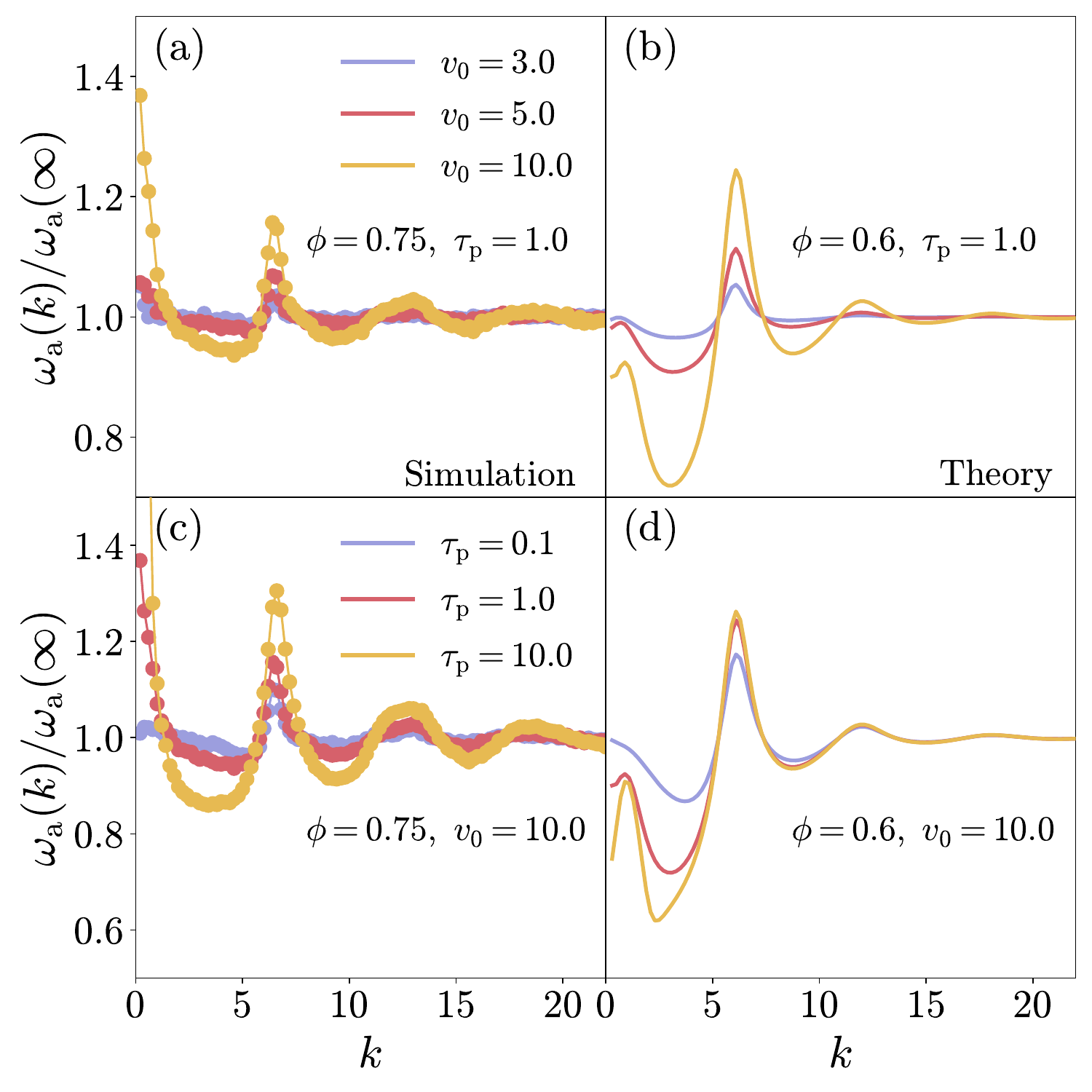}
    \caption{The active-active contribution to the velocity correlation function $\omega_{\mathrm{a}}(k)$ (normalized by $\omega_{\mathrm{a}}(\infty)=v_{0}^{2}/2$) as a function of wavenumber $k$, directly measured from simulation data (a,c) or obtained fully analytically via the ITT formalism (b,d). Results correspond to different active speeds (a-b) and persistence times (c-d).}
    \label{omegak_active_thermal}
\end{figure}

Using~\cref{vv_ffexpression_ITT}, we have calculated $\omega_{\mathrm{a}}(k)$ for different active speeds $v_{0}$ and persistence times $\tau_{\mathrm{p}}$. Results for the same active control parameters have also been retrieved from the simulation data of thermal ABPs and both have been plotted in~\cref{omegak_active_thermal}. In accordance with the athermal simulation results for $\omega(k)$~\cite{FlennerAOUP2016,BerthierAOUP2017,Flenner2020,Szamel_2021}, we witness the emergence of oscillations around $\omega_{\mathrm{a}}(\infty)=v_{0}^{2}/2$ whose relative size increases as either $v_{0}$ or $\tau_{\mathrm{p}}$ is increased. The locations of the corresponding peaks (approximately) coincide with the ones from the static structure factor and reveal a clear structural signature in this velocity correlation. Interestingly, all these qualitative features are thus also predicted by our ITT method with even a reasonable degree of quantitative accuracy. This is quite remarkable as we reiterate that the whole ITT procedure has only required an analytical passive structure factor as input and we only consider the first non-trivial active modes. At the same time, the ITT results fail to capture the correct behavior at small wavenumbers $k$, especially for large persistence and active speed where the value of $\omega_{\mathrm{a}}$ from simulations is seen to increase dramatically, whereas the theoretical results in some cases even bend down again. It is known, however, that MCT-approximations can sometimes yield less accurate results in the small-$k$ limit which already occurs for passive systems~\cite{gotze2008complex}. 

To place our thermal results in a broader context, we have also extracted active-active velocity correlations from the simulation data of athermal ABPs (see~\cref{Sk_omegak_active_athermal}). These exhibit the same qualitative behavior as the corresponding thermal results, though the magnitude of the oscillations is larger indicating that thermal noise disrupts the emergence of these correlations. 
Overall, the obtained results for $\omega_{\mathrm{a}}(k)$ clearly demonstrate the existence of spatial active velocity correlations in thermal and athermal dense active matter (which have been checked for finite size effects). This carries an important implication, as it has been argued for athermal systems that such correlations only arise when one considers the total velocity consisting of both the active and interaction forces~\cite{Henkes2020,Caprini2020}. Our results instead suggest that this is not strictly necessary.

\subsubsection{Active-passive and passive-passive velocity correlations}
In contrast to the active-active correlations, we find that the ITT approach has proven less fruitful in its efforts to describe the other two contributions [$\omega_{\mathrm{c}}(k)$ and $\omega_{\mathrm{int}}(k)$] to the total velocity correlations.This is probably a consequence of the observables not taking the form of density doublets, so an additional projection is needed. For the latter term, the presence of two interaction forces hinders a direct analytical evaluation when using the standard MCT-approximations. As such, an additional projection operator seems to be necessary to separate both interaction forces but this has not yielded satisfying results. Interestingly, by means of a properly orthogonalized projection operator on density doublets one can find qualitatively consistent results for $\omega_{\mathrm{int}}(k)$ of a passive Brownian system (see~\cref{FF_corr_passive} for more details), which corresponds to taking only the first term in~\cref{ITT_ABP}. In comparison, due to the presence of only one interaction force, the active-passive correlation function $\omega_{\mathrm{c}}(k)$ can be directly evaluated. Though its sign and order of magnitude are accurately captured, the qualitative behavior we find is inconsistent with results from simulations (see \cref{sec:vv_analy} for more details). It thus seems that a conventional projection on density doublets is insufficient to fully capture the behavior of the cross term. 

\begin{figure}[ht!]
    \centering
    \includegraphics[width=0.48\textwidth]{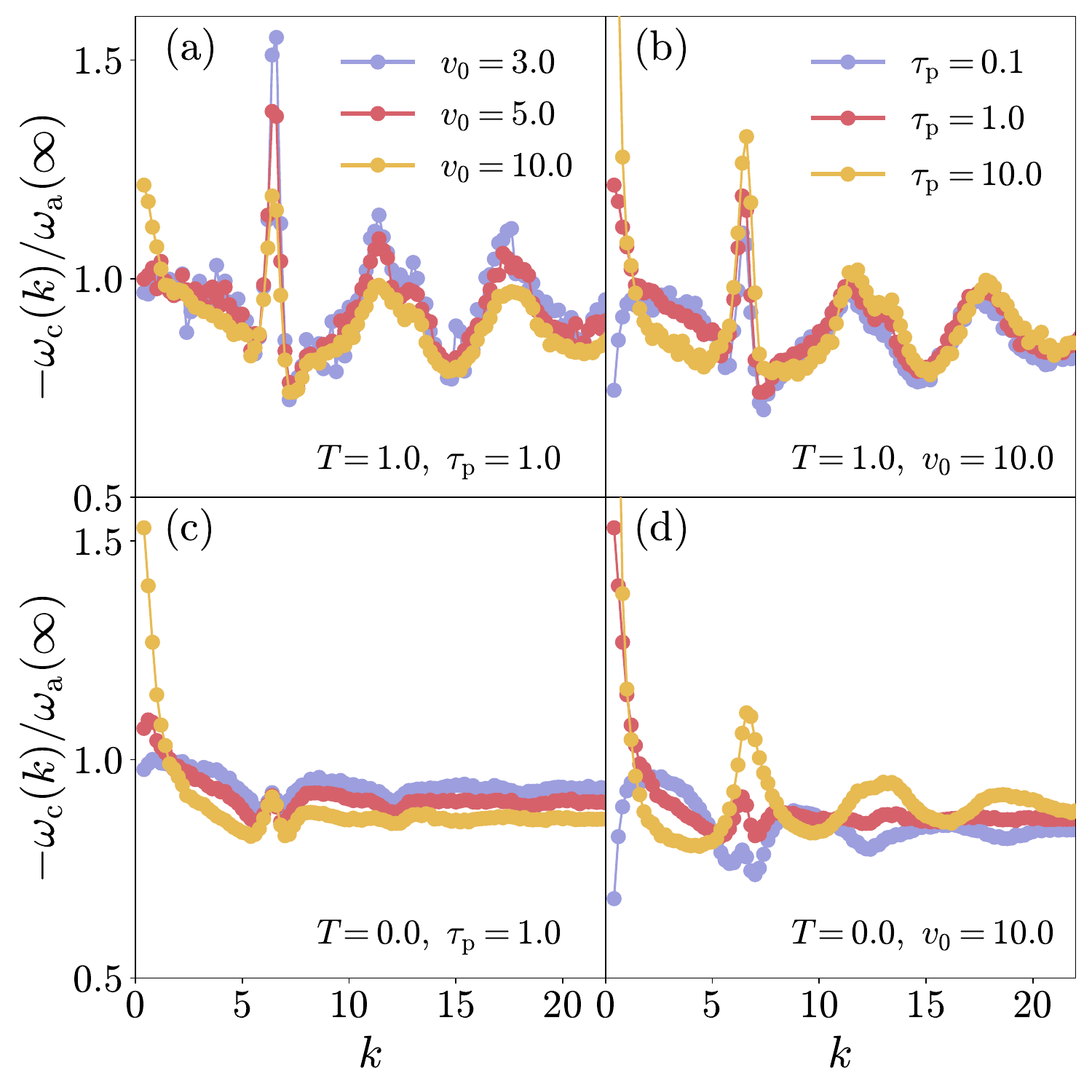}
    \caption{The active-passive contribution to the velocity correlation function $\omega_{\mathrm{c}}(k)$ (normalized by $\omega_{\mathrm{a}}(\infty)=v_{0}^{2}/2$ and with a prefactor of minus one to improve visibility) as a function of wavenumber $k$. The results are directly measured from simulation data for (a,b) a thermal system at temperature $T=1.0$ and (c,d) an athermal system, and correspond to different active speeds (a,c) and persistence times (b,d).}
    \label{omegak_cross_thermal_athermal}
\end{figure}

As such, we primarily focus on the results obtained from both thermal and athermal simulations and try to scrutinize the role of thermal noise. Let us first focus on the active-passive velocity correlations for which the results are shown in~\cref{omegak_cross_thermal_athermal}. Note that since the active-passive velocity correlation function is always negative (implying an anticipated anti-correlation between active and passive forces) and to allow for a good comparison with the other velocity correlations, we have plotted $-\omega_{\mathrm{c}}(k)$ and normalized the results with $\omega_{\mathrm{a}}(\infty)=v_0^{2}/2$. The first thing we observe is that both the thermal and athermal results are mostly of the same order of magnitude as the active-active velocity correlations and the rescaled magnitude of their asymptotic value $-\omega_{\mathrm{c}}(\infty)$ increases with persistence and decreases for enhanced active speed. 

In analogy to the active-active velocity correlations we also see, provided the active speed and persistence are large enough, a significant increase of $-\omega_{\mathrm{c}}(k)$ when $k$ approaches zero that appears to be more pronounced for the athermal systems. Since the small-$k$ regime indicates long-ranged correlations it is sensible that these are stronger for an athermal system. At the same time, it is worth noting that at a small persistence of $\tau_{\mathrm{p}}=0.1$ (and $v_{0}=10$) the value for $-\omega_{\mathrm{c}}(k)$ has already tipped over and decreases as one approaches $k\rightarrow 0$ (in both the thermal and athermal case). It would be interesting to see if this also occurs for the other curves if one probes at smaller values of $k$. Intuitively, there might even exist a relationship between the so-called persistence length $l_{\mathrm{p}}=v_{0}\tau_{\mathrm{p}}$~\cite{Debets2021cage} and the wavenumber where such a tipping takes place. This could explain why we only see it at a relatively small persistence.

In comparison, for larger wavenumbers the similarity between the thermal and athermal active-passive velocity correlations is less obvious with both demonstrating oscillatory behavior (though not always with the same phase) and a sudden peak at roughly the same location as the peak of $S(k)$. Moreover, these effects for larger values of $k$ and thus shorter length scales (on the order of a particle diameter or smaller) are much more evident for the thermal results. They are therefore probably enhanced by the thermal noise inducing larger and more erratic instantaneous repulsive forces. 

\begin{figure}[ht!]
    \centering
    \includegraphics[width=0.48\textwidth]{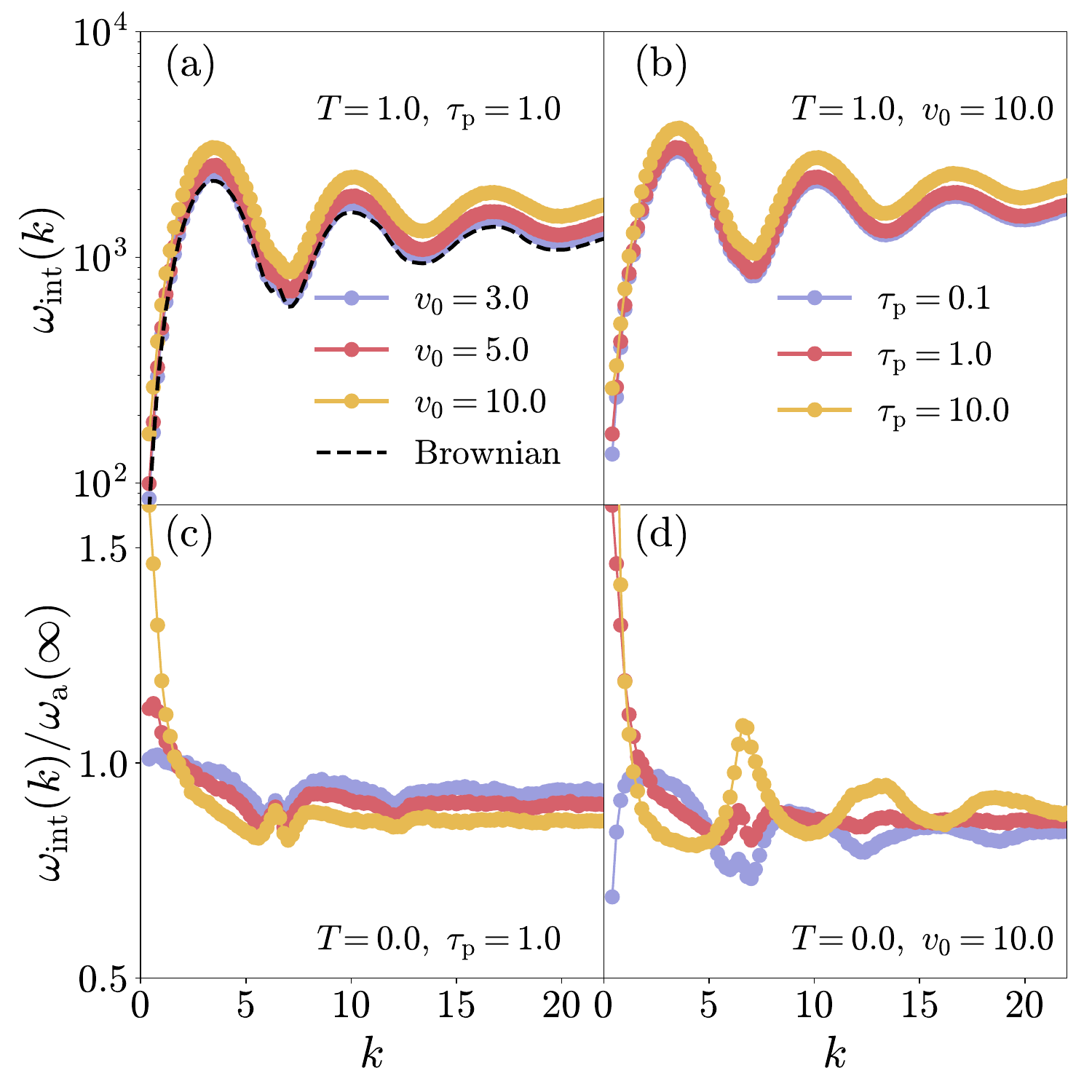}
    \caption{The passive-passive contribution to the velocity correlation function $\omega_{\mathrm{int}}(k)$ as a function of wavenumber $k$, directly measured from simulation data for (a,b) a thermal system at a temperature $T=1.0$ and (c,d) an athermal system (for which it is normalized by $\omega_{\mathrm{a}}(\infty)=v_{0}^{2}/2$). The results correspond to different active speeds (a,c) and persistence times (b,d). The dotted line is obtained from simulations of a passive Brownian system at $T=1.0$.}
    \label{omegak_int_thermal_athermal}
\end{figure}

We finalize our results by examining the passive-passive velocity correlations $\omega_{\mathrm{int}}(k)$ which are shown in~\cref{omegak_int_thermal_athermal}. Upon inspection we may immediately note that the thermal and athermal results are significantly different, both quantitatively and qualitatively. In particular, in the thermal case this term is relatively big and completely dominates over all other contributions to the velocity correlations. We expect this to be caused by translational noise occasionally inducing large instantaneous repulsive forces which greatly exceed the active force. That thermal passive motion is mostly responsible for this correlation is also reflected in the fact that the qualitative behavior and order of magnitude of $\omega_{\mathrm{int}}(k)$ are almost the same as the one obtained from an equivalent Brownian system without self-propulsion (see dotted line in~\cref{omegak_int_thermal_athermal}). Moreover, the oscillations of $\omega_{\mathrm{int}}(k)$ are out of phase with the ones from $S_{\mathrm{neq}}(k)$ and thus the passive-passive velocity correlation function seems anticorrelated with the structure.

The corresponding athermal values are instead much smaller, i.e., of the same order as the active-active and active-passive velocity correlation functions. This can be explained by realizing that in a dense athermal system interaction forces, on average, constantly counteract the self-propulsion forces and are thus expected to be of the same order of magnitude. More strikingly, the results (and thus the qualitative behavior) for $\omega_{\mathrm{int}}(k)$ seem almost indistinguishable from the active-passive velocity correlation function $-\omega_{\mathrm{c}}(k)$. This hints at a possible formal equivalence (or at least strong connection) between these parameters in athermal systems, which is further corroborrated by recent results where it is in fact proved in 1D that $\avg{\mv{F}_{i}\cdot \mv{F}_{i}}_{\mathrm{ss}}=-\avg{\mv{f}_{i}\cdot \mv{F}_{i}}_{\mathrm{ss}}$~\cite{schiltzrouse2023kinetic}.

Overall, when comparing thermal and athermal velocity correlations in dense active fluids it thus seems sensible to focus more on active-active and active-passive correlations. 
Alternatively, coarse-graining the total velocity over a cleverly chosen time window might also mitigate the effects of thermal noise on the velocity correlations. This can be straightforwardly done in simulations, but is likely more tedious in a microscopic description such as ITT. 


\section{Conclusion}
In this work we have sought to characterize nonequilibrium (or steady-state) structure and velocity correlations in dense active systems, with a prime focus on the often overlooked influence of thermal (or translational) noise.
We have done so by means of both particle-based computer simulations and a microscopic theory based on the integration through transients (ITT) formalism in conjunction with active mode-coupling theory. 
Consistent with literature, we find that for both thermal and athermal systems enhanced active speed and persistence diminish the structure and make the system more compressible (the latter being manifested by an increase of the nonequilibrium static structure factor in the limit of zero wavenumber). More importantly, we also demonstrate that all these features can be qualitatively predicted by our microscopic theory which only takes a passive (analytical) static structure factor as its input. 

Besides structure, we have also examined all distinct contributions to the equal-time velocity correlations, i.e., spatial correlations between active velocities, interaction forces, and cross terms of one with the other. We show that our theory is equally capable of making accurate qualitative predictions for the active-active velocity correlations. These correlations are similar in both thermal and athermal active systems (though weaker for the former) and become more significant (larger oscillations and a stronger increase at small wavenumber) upon increasing the active speed or persistence. The existence of such correlations is especially interesting considering that for athermal systems it has been previously argued that spatial velocity correlations only arise when one takes into account the total velocity consisting of both the active and interaction force.

Moreover, we demonstrate that the cross correlation exhibits distinct qualitative changes due to thermal noise but only when one probes length scales on the order of a particle diameter or smaller. The correlation between interaction forces instead is completely dominated by thermal noise which leads to much larger values and different qualitative behavior on all length scales.   

Overall, our results further establish ITT as a promising route to evaluate nonequilibrium averages in active matter and suggest that correlations between active velocities might carry more information than previously anticipated. They also show that thermal noise strongly influences velocity correlations that involve the instantaneous interaction force. A possible remedy for this could be to coarse-grain the total velocity over an astutely chosen time window, which could be done straightforwardly in simulations, less so in a microscopic description such as ITT. 

As a followup, it would be highly interesting to study the small-wavenumber behavior of these individual contributions to the velocity correlations in greater detail and possibly extract scaling relations that can be compared for athermal and thermal systems~\cite{Szamel_2021}. This, however, requires (especially for thermal ABPs) extensive simulation efforts and is therefore left for future work.


\begin{acknowledgments}

The authors acknowledge financial support from the Dutch Research Council (NWO) through a START-UP grant (LMCJ and VED) and from the European Union through an ERASMUS+ grant (LS).

\end{acknowledgments}

\appendix

\section{Symmetry properties}\label{sec:real_ISF}
In this appendix we formally show, basing ourselves on the derivation of the symmetry properties of the ISF in Ref.~\cite{Reichert_20_Thesis}, how the inclusion of a factor $i^{l}$ in our density mode $\rho_{l}(\mv{k})$ in combination with choosing the wavevector $\mv{k}$ along the $x$-axis ensures that the transient density correlations are always real.
It has been shown that if we rotate $\boldsymbol{k}$ by an angle $\delta\theta$, the transient density correlation is modified in the following way~\cite{Reichert_20_Thesis}
\begin{equation}
    S_{ll'}(\boldsymbol{k}_{+\delta\theta},t) = e^{-i(l-l')\delta\theta}S_{ll'}(\boldsymbol{k},t)
\end{equation}
In particular, taking $\delta\theta=\pi$, one finds 
\begin{equation}\label{complex_1}
S_{ll'}^*(\boldsymbol{k},t) = (-1)^{l-l'}S_{-l,-l'}(\boldsymbol{k},t),
\end{equation}
while if we let $\delta\theta=-\theta_{\boldsymbol{k}}$, we have
\begin{equation}\label{Stilde_def}
\tilde{S}_{ll'}(k,t)\equiv S_{ll'}(k\boldsymbol{e_x},t) = e^{i(l-l')\theta_{\boldsymbol{k}}}S_{ll'}(\boldsymbol{k},t).
\end{equation}
Next, let us define the linear transformation $\mathcal{T}$ 
\begin{equation}
    \left(\begin{tabular}{c}
         $r_x$ \\
         $r_y$\\
         $\theta$
    \end{tabular}\right)
    \rightarrow
    \left(\begin{tabular}{c}
         $r_x$ \\
         $-r_y$\\
         $-\theta$
    \end{tabular}\right),
\end{equation}
which represents a symmetric transformation with respect to the $x$-axis, and thus leaves absolute distances invariant. Therefore, the interaction potential $U$ is unchanged, meaning $P_{\mathrm{eq}}$ is invariant under $\mathcal{T}$. It is straightforward to prove that $\Omega^\dagger$ is also invariant. This implies that $S_{ll'}(k\boldsymbol{e}_{x},t)=(-1)^{l-l'}S_{-l,-l'}(k\boldsymbol{e}_{x},t)$, which combined with \cref{complex_1} yields
\begin{equation}\label{Stilde_sym}
    \tilde{S}_{ll'}(k,t) =(-1)^{l-l'}\tilde{S}_{-l,-l'}(k,t) = \tilde{S}^*_{ll'}(k,t),
\end{equation}
and proves that $\tilde{S}_{ll^\prime}(k,t)$ is always real.

\section{Analytical details}\label{sec:analy_details}
In this appendix we will present more detailed derivations of several analytical expressions shown throughout the main text.
\subsection{Correlation of swim velocity corrections}
The irreducible correlation of swim velocity corrections, i.e., $C_{\text{irr}}^v (t)$, serves to normalize the ITT quantities we compute [see \cref{CCirr_def}].
To find an analytical expression for this term we employ standard MCT-approximations. In particular, we insert two projections onto density doublets,
\begin{equation}
    \mathcal{P}_{2} = \frac{1}{2}\sum_{\mv{q}_{1}\mv{q}_{2}}\sum_{l_{1}l_{2}}   |\rho_{l_{1}}(\mv{q}_{1}) \rho_{l_{2}}(\mv{q}_{2})\big\rangle S^{-1}_{l_{1}l_{1}}(q_{1})S^{-1}_{l_{2}l_{2}}(q_{2}) 
      \big\langle \rho^{*}_{l_{1}}(\mv{q}_{1})\rho^{*}_{l_{2}}(\mv{q}_{2})|,
\end{equation}
such that
\begin{equation}
    C_{\text{irr}}^v (t)\approx \avg{\sum\limits_{j=1}^N \boldsymbol{e}_j\cdot\boldsymbol{F}_j\ \mathcal{P}_{2} e^{\Omega_{\text{irr}}^\dagger t} \mathcal{P}_{2}\ \sum\limits_{i=1}^N \boldsymbol{e}_i\cdot\boldsymbol{F}_i}.
\end{equation}
Note that, in contrast to conventional MCT, one could have also used a projection on density singlets but it can be shown that these yield vanishing contributions.  Indeed, if we consider $\moy{\boldsymbol{e}_j\cdot\boldsymbol{F}_j \rho_l(\boldsymbol{k})}$, then $\boldsymbol{k} = \boldsymbol{0}$ by translational invariance. Since $P_{\mathrm{eq}}$ is independent of orientation angles, we can isolate the integral over positions from the one over orientations, leaving us with $\moy{\boldsymbol{e}_j\rho_l(\boldsymbol{0})}_{\{\theta_i\}}\cdot\moy{\boldsymbol{F}_j}_{\{\boldsymbol{r}_i\}}$. The second term is zero because $\boldsymbol{F}_j P_{\mathrm{eq}}\propto\nabla_jP_{\mathrm{eq}}$ and the equilibrium probability vanishes at infinity. As a result the singlet projection will not contribute and we therefore use the next-leading doublet projection.
For the so-called left vertex we then have (as has been previously calculated in Ref.~\cite{Reichert_20_Thesis})
\begin{equation}\label{left_vertex_ITT}
\begin{aligned}
    &V_{L} \equiv\frac{1}{2}\sum\limits_{j=1}^N \moy{\boldsymbol{e}_j\cdot\boldsymbol{F}_j \rho_{l_1}(\boldsymbol{q}_{1}) \rho_{l_2}(\boldsymbol{q}_{2})} S_{l_1l_1}^{-1}(q_{1})S_{l_2l_2}^{-1}(q_{2})\\
    &= -\frac{\rho(l_1+l_2)}{4\beta} \delta_{\boldsymbol{q}_1,-\boldsymbol{q}_2}\delta_{|l_1+l_2|,1} q_1 c(q_1)\argp{e^{il_2 \theta_{q_1}}\delta_{l_1,0} - e^{il_1 \theta_{q_1}}\delta_{l_2,0}},
\end{aligned}\end{equation}
where we have used partial integration to rewrite the interaction force. We emphasize that the left vertex also naturally arises when applying MCT-approximations to
$C_{\mathrm{irr}}^A (t)$ (for any observable $A$). Moreover, due to the symmetry of $C^{v}_{\mathrm{irr}}(t)$, one can show that the right vertex is the complex conjugate of the left one. The other MCT-approximation consists of factorizing the four-point density correlation function and replacing irreducible with full dynamics, i.e.,
\begin{equation}\label{Fact_approx}
\begin{split}
    & \avg{\rho^{*}_{l_{1}}(\mv{q}_{1})\rho^{*}_{l_{2}}(\mv{q}_{2}) e^{\Omega_{\mathrm{irr}}^{\dagger}t} \rho_{l_{3}}(\mv{q}_{3}) \rho_{l_{4}}(\mv{q}_{4}) } \approx S_{l_{1}l_{3}}(\mv{q}_{1},t) \\[3pt]
    & S_{l_{2}l_{4}}(\mv{q}_{2},t)\ \delta_{\mv{q}_{1},\mv{q}_{3}}\delta_{\mv{q}_{2},\mv{q}_{4}} + S_{l_{1}l_{4}}(\mv{q}_{1},t)S_{l_{2}l_{3}}(\mv{q}_{2},t)\ \delta_{\mv{q}_{1},\mv{q}_{4}}\delta_{\mv{q}_{2},\mv{q}_{3}}.
\end{split}
\end{equation}
Combining these results and taking the thermodynamic limit one finally arrives at
\begin{equation}
\begin{aligned}
   & \frac{D_t\beta^2}{N}C_{\text{irr}}^v(t)= \frac{\rho D_t}{8\pi}  \int_0^\infty dq\, q^3 c(q)^2 \sum\limits_{\lambda,\lambda' = \pm 1}\lambda\lambda'\\
   &\qquad \qquad \times\argp{\tilde{S}_{\lambda\lambda'}(q,t)\tilde{S}_{00}(q,t) + \tilde{S}_{0\lambda'}(q,t)\tilde{S}_{\lambda0}(q,t)},
    \end{aligned}
\end{equation}
which is consistent with the expression shown in Refs.~\cite{Reichert2021rev,Reichert_20_Thesis}.

\subsection{Nonequilibrium static structure factor}
For the static structure factor we take $A=\rho^*_{0}(\mv{k})\rho_{0}(\mv{k})$. The reference contribution to the steady-state average is thus simply $\avg{A}=S(k)$. For the integral term we again use a projection on density doublets, but since $A$ is already of the form of a density doublet we only require one projection, i.e.,
\begin{equation}
    C^{A}_{irr}(t)\approx \avg{\sum\limits_{j=1}^N \boldsymbol{e}_j\cdot\boldsymbol{F}_j\ \mathcal{P}_{2} e^{\Omega_{\text{irr}}^\dagger t} \rho^*_{0}(\mv{k})\rho_{0}(\mv{k})}
\end{equation}
Invoking the expression for the left vertex [\cref{left_vertex_ITT}], the factorization approximation [\cref{Fact_approx}], and the symmetry properties of $\tilde{S}_{ll^\prime}(k,t)$ [\cref{Stilde_sym}] one can then find that 
\begin{equation}
    C^{A}_{irr}(t)\approx 2\rho \beta^{-1} k c(k) \tilde{S}_{10}(k,t) \tilde{S}_{00}(k,t)
\end{equation}
which leads to~\cref{S_ITTexpression} in the main text.

\subsection{Active-active velocity correlations}\label{omega_a_der}
To derive an expression for the active-active velocity correlations we start by noticing that the product of an active force and the zeroth density mode can be rewritten as a combination of higher-order density modes, i.e.,
\begin{equation}\label{active_densitymode}
\frac{1}{\sqrt{N}\zeta}\sum\limits_{j=1}^{N} \hat{\boldsymbol{k}}\cdot\boldsymbol{f}_j e^{i\boldsymbol{K}\cdot\boldsymbol{r}_j} = \frac{v_0}{2}\sum\limits_{\epsilon=\pm 1} i^\epsilon e^{i\epsilon\theta_{\boldsymbol{k}}}\rho_{-\epsilon}(\boldsymbol{K}) 
\end{equation}
Thus, we have for the reference contribution to the active-active velocity correlation
\begin{equation}\label{eq_vv_ff}
\begin{aligned}
    \avg{A}&=\frac{1}{N\zeta^2} \moy{\sum\limits_{i=1}^{N} \hat{\boldsymbol{k}}\cdot\boldsymbol{f}_i e^{-i\boldsymbol{k}\cdot\boldsymbol{r}_i}\sum\limits_{j=1}^{N} \boldsymbol{f}_j\cdot\hat{\boldsymbol{k}}  e^{i\boldsymbol{k}\cdot\boldsymbol{r}_j}}\\
    & = \frac{v_0^2}{4}(S_{11}(k) + S_{-1-1}(k)) = \frac{v_0^2}{2},
\end{aligned}\end{equation}
while for the contribution inside the integral we again only require one projection on density doublets,
\begin{equation}
    C^{A}_{\mathrm{irr}}(t)\approx \frac{1}{N\zeta^2}\avg{\sum\limits_{m=1}^N \boldsymbol{e}_m\cdot\boldsymbol{F}_m\ \mathcal{P}_{2} e^{\Omega_{\text{irr}}^\dagger t} \sum\limits_{i,j} \hat{\boldsymbol{k}}\cdot\boldsymbol{f}_i e^{-i\boldsymbol{k}\cdot\boldsymbol{r}_i} \boldsymbol{f}_j\cdot\hat{\boldsymbol{k}}  e^{i\boldsymbol{k}\cdot\boldsymbol{r}_j}}
\end{equation}
Invoking the expression for the left vertex [\cref{left_vertex_ITT}], the factorization approximation [\cref{Fact_approx}], and the symmetry properties of $\tilde{S}_{ll^\prime}(k,t)$ [\cref{Stilde_sym}] one may find
\begin{equation}\label{ITTcorr_vv_ff}
      C_{\text{irr}}^{A}(\boldsymbol{k},t) = v_0^2\beta^{-1} \rho k c(k)  \tilde{S}_{01}(k,t)[\tilde{S}_{11}(k,t) - \tilde{S}_{-11}(k,t)].
\end{equation}
Using these results in combination with \cref{ITT_ABP,CCirr_def} then yields~\cref{vv_ffexpression_ITT} in the main text.



\subsection{Active-passive velocity correlations}\label{sec:vv_analy}
Despite not giving fully satisfying results, one can find (with standard MCT-approximations) an explicit expression for the active-passive velocity correlations. 
Let us first recall the definition of this correlation,
\begin{equation}
    \omega_{\mathrm{c}}(k) = \frac{1}{2N\zeta^2} \sum\limits_{\boldsymbol{K}=\pm\boldsymbol{k}}\moy{\sum\limits_{i,j=1}^{N} \hat{\boldsymbol{k}}\cdot \boldsymbol{F}_j\,\hat{\boldsymbol{k}}\cdot\boldsymbol{f}_i e^{i\boldsymbol{K}\cdot (\boldsymbol{r}_j-\boldsymbol{r}_i)}}_{\mathrm{ss}}
\end{equation}
where we mention that one can also compute the two cross-correlations separately then sum them, but this gives the same result as presented below. We choose to evaluate the sum directly as it allows for simplifications much earlier in the computation.

Because $P_{\mathrm{eq}}$ does not depend on the angles $\theta_{i}$, the reference contribution to the active-passive velocity correlations is simply zero, i.e., $\avg{A}=0$. For the contribution inside the integral we now require two projections on density doublets. This introduces a right vertex of the form
\begin{equation}
    \frac{1}{2N\zeta^2}\moy{\rho_{l_3}^*(\boldsymbol{q}_3)\rho_{l_4}^*(\boldsymbol{q}_4)\sum\limits_{i,j=1}^{N} \hat{\boldsymbol{k}}\cdot\boldsymbol{F}_j\,\hat{\boldsymbol{k}}\cdot\boldsymbol{f}_i e^{i\boldsymbol{K}\cdot(\boldsymbol{r}_j-\boldsymbol{r}_i)}}.
\end{equation}
Performing integration by parts to remove the interaction force and summing over $\boldsymbol{K}=\pm\boldsymbol{k}$, we end up with two types of terms, which contain either averages of three ($S_a$) or four ($S_{b}$) density modes. For the latter, we can use a Gaussian approximation, which immediately fixes $(\boldsymbol{q}_3,l_3)$ and $(\boldsymbol{q}_4,l_4)$ to the pair $(-\boldsymbol{K},-\epsilon),(\boldsymbol{K},0)$ (or vice versa), and for the former, we can use the convolution approximation~\cite{Voigtmann2017}, which will only fix $(\boldsymbol{q}_3+\boldsymbol{q}_4,l_3+l_4)$ to $(\boldsymbol{0},-\epsilon)$, and leave one degree of freedom in momentum and in angular mode.

One can then show that the two respective contributions yield for the product of the four-point dynamic density correlation with the right vertex the following expressions,
\begin{equation}
    \begin{aligned}        
    & S_a = \frac{D_t v_0}{4N}\sum\limits_{\epsilon =\pm 1} \epsilon e^{i\epsilon\theta_{\boldsymbol{k}}}\sum\limits_{\mv{q}_{3},l_{3}} (-\hat{\boldsymbol{k}}\cdot\boldsymbol{q}_3) \\
     &\qquad \times \left<\rho^*_1\rho^*_2 e^{\Omega_{\text{irr}}^\dagger t}\rho_{l_3}(\boldsymbol{q}_3)\rho_{-l_3-\epsilon}(-\boldsymbol{q}_3)\right> \\
    &\qquad \times S_{l_3l_3}^{-1}(q_3)\big[S_{l_3l_3}(|\boldsymbol{k}-\boldsymbol{q}_3|) +  S_{l_3l_3}(|\boldsymbol{k}+\boldsymbol{q}_3|)\big] \\[4pt]
    &S_b = \frac{D_t v_0}{4}k\sum\limits_{\epsilon =\pm 1} \epsilon e^{i\epsilon\theta_{\boldsymbol{k}}}\left[\left<\rho^*_1\rho^*_2 e^{\Omega_{\text{irr}}^\dagger t}\rho_{-\epsilon}(-\boldsymbol{k})\rho_{0}(\boldsymbol{k})\right>\right. \\
    &\qquad \left.-\left<\rho^*_1\rho^*_2 e^{\Omega_{\text{irr}}^\dagger t}\rho_{-\epsilon}(\boldsymbol{k})\rho_{0}(-\boldsymbol{k})\right>\right],
    \end{aligned}
\end{equation}
where $\rho_1 \equiv \rho_{l_1}(\boldsymbol{q}_1)$ is defined for convenience. Finally, introducing the left vertex [\cref{left_vertex_ITT}], the factorization approximation [\cref{Fact_approx}], and the symmetry properties of $\tilde{S}_{ll^\prime}(k,t)$ [\cref{Stilde_sym}] we find for the total contribution of each term (after a change of variables),
\begin{equation}\label{ITTcorr_vv_fF}
\begin{aligned}
   &  C_{\text{irr}}^{A,a}(\boldsymbol{k},t) = \frac{D_t v_0 }{8\pi^2\beta} \int d\boldsymbol{q} q^2 \,c(q)  \cos(\theta_{\boldsymbol{k}} - \theta_{\boldsymbol{q}})^2 \\
   &\qquad \times \argp{1 - S(|\boldsymbol{k}-\boldsymbol{q}|)S^{-1}(q)} \Big(\tilde{S}_{0,0}(q,t)\tilde{S}_{1,1}(q,t)\\
   & \qquad  - \tilde{S}_{0,0}(q,t)\tilde{S}_{1,-1}(q,t) + 2\tilde{S}_{01}(q,t)\tilde{S}_{10}(q,t)\Big)\\
    & C_{\text{irr}}^{A,b}(\boldsymbol{k},t) = \frac{ D_t v_0}{2\beta} \rho c(k) k^2 \\
    & \times  \left(\tilde{S}_{0,0}(k,t)(\tilde{S}_{1,1}(k,t)-\tilde{S}_{1,-1}(k,t)) + 2\tilde{S}_{0,1}(k,t)\tilde{S}_{1,0}(k,t)\right),
\end{aligned}  
\end{equation}
where $C^{A}_{\mathrm{irr}}=C^{A,a}_{\mathrm{irr}}+C^{A,b}_{\mathrm{irr}}$. With the help of the above result one can calculate the active-passive velocity correlation function but this turns out to give qualitatively inconsistent results, though the negative sign and order of magnitude are correctly captured. It is therefore likely that the second projection on density doublets introduces a new error and requires refinement to allow for a better prediction. 

\section{Additional data for athermal systems}
\begin{figure}[ht!]
    \centering
    \includegraphics[width=0.48\textwidth]{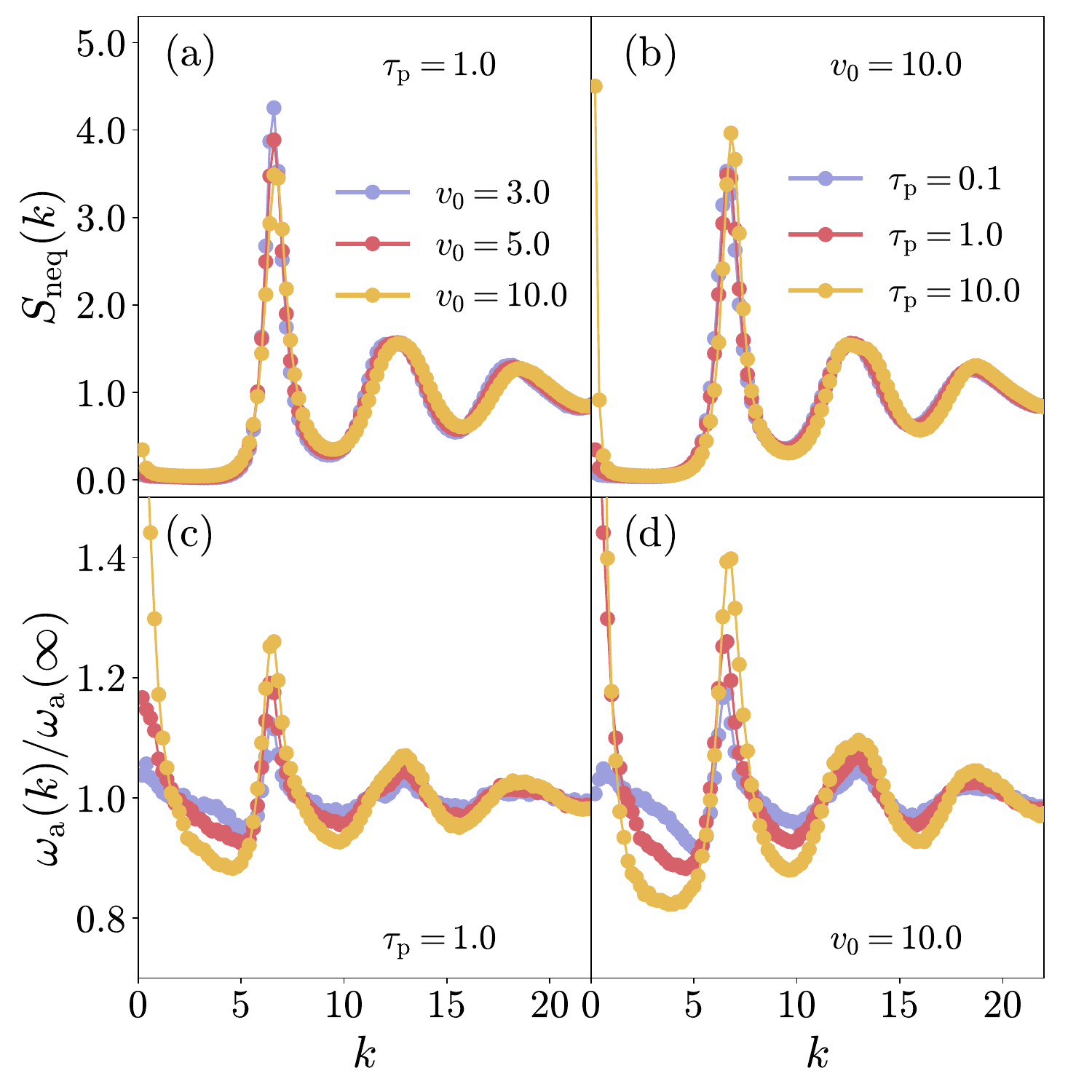}
    \caption{(a,b) The static structure factor and (c,d) active-active contribution to the velocity correlation function $\omega_{\mathrm{a}}(k)$ (normalized by $\omega_{\mathrm{a}}(\infty)=v_{0}^{2}/2$) as a function of wavenumber $k$, directly measured from simulation data for athermal system. The results correspond to different active speeds (a,c) and persistence times (b,d).}
    \label{Sk_omegak_active_athermal}
\end{figure}

In \cref{Sk_omegak_active_athermal} we have plotted results for $S_{\mathrm{neq}}(k)$ and $\omega_{\mathrm{a}}(k)$ obtained from simulations of athermal ABPs. Overall, these show similar behavior as their thermal counterparts with one notable exception, i.e., $v_{0}=10.0$, $\tau_{\mathrm{p}}=10.0$, where the peak of $S_{\mathrm{neq}}(k)$ is seen to increase again. This, as well as the very steep rise at vanishing wavenumber, can be explained by realizing that for such large values of the active control parameters the system has undergone MIPS. We mention that the nonmonotonic behavior of the first peak height has also been reported in previous work~\cite{Biniossek2018}. 

\section{Passive-passive velocity correlations for a Brownian system}\label{FF_corr_passive}
As mentioned in the main text, the passive-passive velocity correlations $\omega_{\mathrm{int}}(k)$ cannot be directly evaluated when employing the standard set of MCT-approximations. This is primarily due to the presence of two interaction forces which necessitates the use of at least one additional projection operator. Adding such a projection makes the evaluation of $C^{A}_{\mathrm{irr}}(t)$ rather intractable (or requires even more approximations). If we instead only focus on a passive system (for which the integral term in~\cref{ITT_ABP} is simply zero), analytical progress can in fact be made.

We thus seek to calculate the passive-passive velocity correlations for a Brownian system which are defined as,
\begin{equation}\label{eq_FF_def}
    \omega^{\mathrm{eq}}_{\mathrm{int}}(k)=\frac{1}{N\zeta^2} \moy{\sum\limits_{i=1}^{N} \hat{\boldsymbol{k}}\cdot\boldsymbol{F}_i e^{-i\boldsymbol{k}\cdot\boldsymbol{r}_i}\sum\limits_{j=1}^{N} \boldsymbol{F}_j\cdot\hat{\boldsymbol{k}}  e^{i\boldsymbol{k}\cdot\boldsymbol{r}_j}}.
\end{equation}
Assuming that forces are primarily mediated via direct interactions between two particles, we introduce the following orthogonalized projection on density doublets for a non-trivial sum of their wavevectors (i.e., $\mv{q}_{1}+\mv{q}_{2} = \mv{k}$)
\begin{equation}\label{def_P2_ortho}
\begin{aligned}
    \projP_2^\perp = & \sum_{\substack{\boldsymbol{q}_1,\boldsymbol{q}_2}} 
    \argp{\left|\rho_1\rho_2\right> - \frac{1}{\sqrt{N}}S_1 S_2\left|\rho_{1+2}\right>} \\
    & \times \frac{S_1^{-1}S_2^{-1}}{2}
    \argp{\left<\rho_1^*\rho_2^*\right| - \frac{1}{\sqrt{N}}S_1 S_2\left<\rho_{1+2}^*\right|}\\
\end{aligned}
\end{equation}
where we have defined $\rho_1 \equiv \rho(\boldsymbol{q}_1)$ and $S_1\equiv S(q_1)$ for convenience (note that the angular indices are dropped entirely since our system is passive). Before proceeding we emphasize that orthogonalizing the projection operator is crucial to obtain meaningful results. Using this projection we can then approximate
\begin{equation}
    \omega^{\mathrm{eq}}_{\mathrm{int}}(k)\approx \frac{1}{N\zeta^2} \moy{\sum\limits_{i=1}^{N} \hat{\boldsymbol{k}}\cdot\boldsymbol{F}_i e^{-i\boldsymbol{k}\cdot\boldsymbol{r}_i}\ \mathcal{P}^{\perp}_{2}\ \sum\limits_{j=1}^{N} \boldsymbol{F}_j\cdot\hat{\boldsymbol{k}}  e^{i\boldsymbol{k}\cdot\boldsymbol{r}_j}},
\end{equation}
which in turn can be evaluated to give (after taking the thermodynamic limit)
\begin{equation}\label{eq_vv_FF}
\begin{aligned}
   &\omega^{\mathrm{eq}}_{\mathrm{int}} (k) = \frac{D_t^2 \rho}{8\pi^2}\int d\boldsymbol{q} S(q) S(|\boldsymbol{k} -\boldsymbol{q}|) \\
   &\qquad \qquad \times \argc{\hat{\boldsymbol{k}}\cdot\boldsymbol{q} \,c(q) + \hat{\boldsymbol{k}}\cdot(\boldsymbol{k} -\boldsymbol{q}) \,c(|\boldsymbol{k} -\boldsymbol{q}|)}^2,
\end{aligned}\end{equation}
and only depends on the structure factor $S(k)$ (and the relevant control parameters).

\begin{figure}[ht!]
    \centering
    \includegraphics[width=0.48\textwidth]{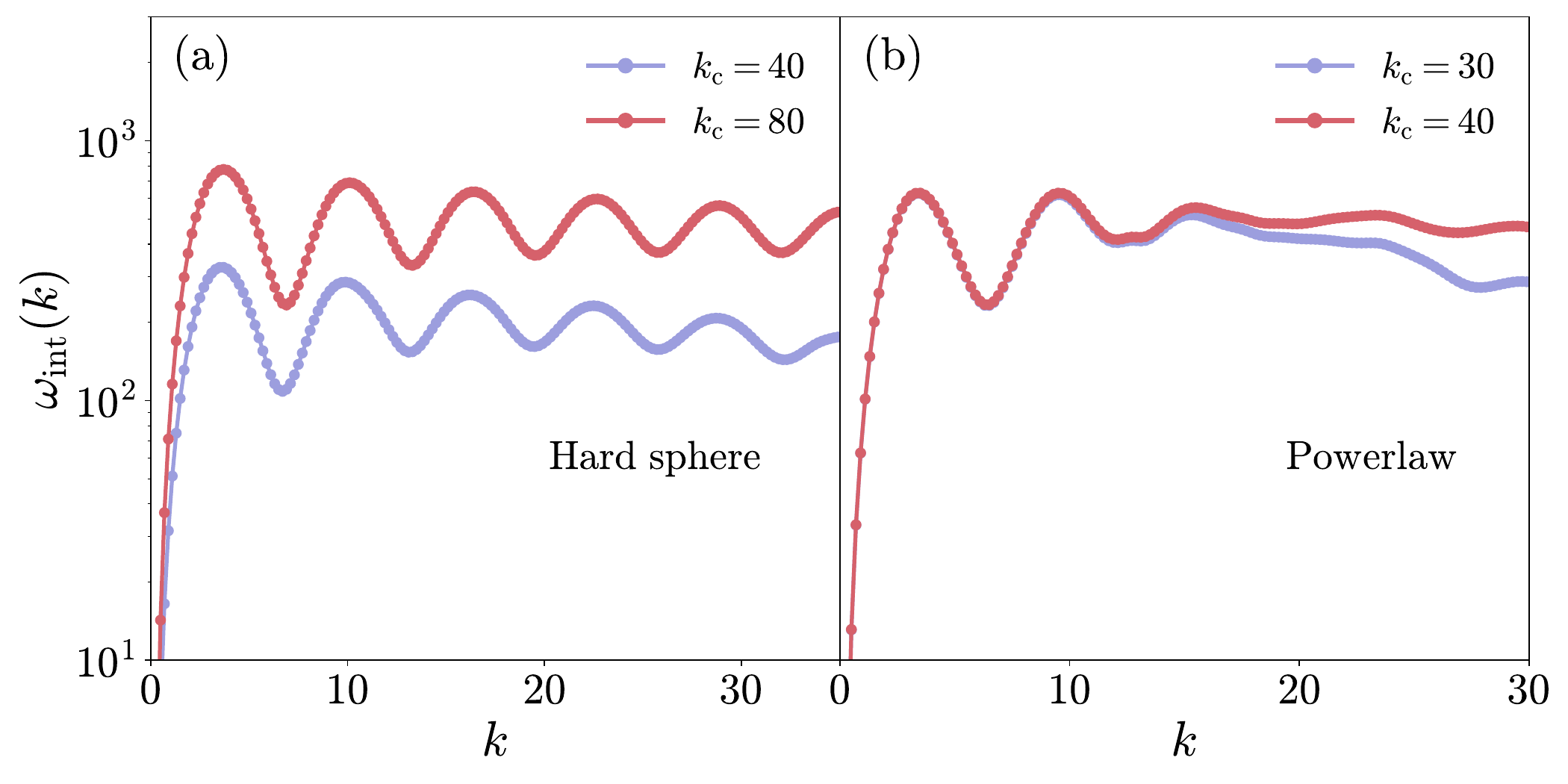}
    \caption{The passive-passive contribution to the velocity correlation function $\omega_{\mathrm{int}}(k)$ as a function of wavenumber $k$. The results are obtained by employing an orthogonalized projection on density doublets using (a) an analytical structure factor for monodisperse passive hard spheres at packing fraction $\phi=0.6$ and (b) a structure factor measured from simulation data of a passive system at temperature $T=1.0$ and packing fraction $\phi=0.75$. The projection introduces an integral over $k$ for which different cutoffs $k_{\mathrm{c}}$ have been chosen.}
    \label{omegak_int_passive}
\end{figure}

\Cref{eq_vv_FF} can be numerically solved employing a standard scheme in MCT where the wavevector integral is rewritten in terms of $q$ and $p=\abs{\mv{k}-\mv{q}}$~\cite{Bayer2007}. We have extracted the solutions to this equation using structure factors obtained from density functional theory~\cite{Thorneywork2018} ($\phi=0.6$) and from our simulations with $v_{0}=0$. The results are plotted for different wavenumber cutoffs $k_{\mathrm{c}}$ in~\cref{omegak_int_passive}. Interestingly, it can be seen that the qualitative behavior is fully consistent with the results obtained directly from simulations (see dotted line in~\cref{omegak_int_thermal_athermal}). We may also note that the results start to fall off for large wavenumbers if one chooses a cutoff that is too small. Finally, increasing the cutoff radius shifts the curves based on the analytical $S(k)$ whereas it is seen to have converged at small $k$ for the simulation $S(k)$. This is likely a result of the hard sphere nature which should lead to diverging forces if one probes small enough length scales, that is, large enough $k$~\cite{Gotze2000HS}.

\bibliography{all}
\end{document}